# Concerning the feasibility of example-driven modelling techniques

S. Thorne, D. Ball, Z. Lawson
UWIC and Cardiff University
SThorne@uwic.ac.uk, DBall@uwic.ac.uk

## 1.0  Introduction

We report on a series of experiments concerning the feasibility of example driven modelling. The main aim was to establish experimentally within an academic environment; the relationship between error and task complexity using a) Traditional spreadsheet modelling, b) example driven techniques. We report on the experimental design, sampling, research methods and the tasks set for both control and treatment groups. Analysis of the completed tasks allows comparison of several different variables. The experimental results compare the performance indicators for the treatment and control groups by comparing accuracy, experience, training, confidence measures, perceived difficulty and perceived completeness. The various results are thoroughly tested for statistical significance using: the Chi squared test, Fisher's exact test for significance, Cochran's Q test and McNemar's test on difficulty.

## 1.1 Example-Driven Modelling

The principle concept of Example Driven Modelling (EDM) is to collect example attribute classifications, provided by the user, to compute the mathematical function of those examples and construct a generalised model via a machine learning technique.

To clarify, figure 1 shows the concept from start to end. Firstly the user would have to provide example attribute classifications for the problem they wish to model. The examples are then formatted into a data set and fed through a learning algorithm. The algorithm learns from the example data, provided which results in a general model, which is able to generalise to new unseen examples in the problem domain.

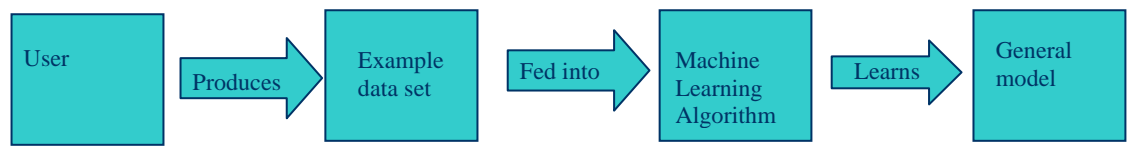

**Figure 1 Example-Driven Modelling (EDM)**

This approach eliminates the need for the user to produce formulae, the user only gives example data for the problem they wish to model. This therefore eliminates errors in constructing formulae since the user is no longer required to produce them.

The burden of calculation is placed on the computer, which using a machine learning algorithm, computes the function of the examples. As the literature suggests, this may be a more effective use of human and computer strengths (Michie, 1989)

However in the case of example giving for EDM this is only theory and some investigation into the feasibility of such an approach is required, i.e. how feasible is it for humans to think up examples for a given problem.

## *2.0 Investigating the feasibility of giving examples*

To investigate if giving examples works in practice an experiment was designed to compare traditional spreadsheet modelling techniques and the novel approach of giving examples. The first group, the "treatment" group, were required to give example data to complete the tasks. The other group, the control group, were given the same tasks to complete using a spreadsheet application.

**2.1 Experimentation**

The experiment into feasibility was designed in accordance guidelines cited by Shadish *et al.* (2002) and Campbell and Stanley (1963). Also, published work using experimental methodologies in spreadsheet research were considered (Hicks and Panko 1995, Javrin and Morrison 1996, Panko and Halverson 1998, Javrin and Morrison 2000, Howe and Simkin 2006)

## **2.2 Experiment aim**

The main aim of the experiment was to establish experimentally within an academic environment, using postgraduate students:

1. The relationship between error and task complexity using a) spreadsheet modelling techniques, b) example giving
2. The (hypothesised) superiority of Example giving over traditional spreadsheet modelling.
3. A satisfactory statistical measure of overconfidence.
4. The relationship between previous spreadsheet experience and accuracy for both traditional spreadsheet modelling and example giving

From these aims and objectives, we will be able to determine the feasibility of Example giving via three performance indicators

1. Whether the participants understand the instruction of giving examples, i.e. can users understand the instructions of giving examples and generate valid examples in the context of the experiment tasks.
2. The accuracy of the examples provided by the participants, i.e. what is the error rate for examples provided by participants

3. The comparative error rate when compared to traditional modelling, i.e. how does the error rate compare to that of traditional modelling and does this warrant further investigation.

## 2.3 Experimental design

The experimental model chosen to evaluate the aims of the experiment is the "Randomised two-group no posttest design". Figure 2 shows the standard design of such experiments, this diagram is read from left to right and shows the

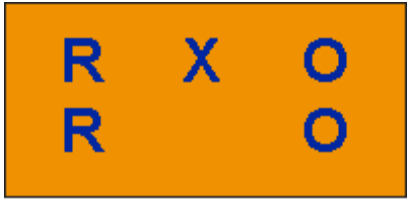

**Figure 2 Randomised two group no post test (Shadish *et al.* 2002)**

The diagram shows the two randomised (R) groups, the treatment group (X), the control group (which is left blank) and the two outcomes (O). In this case the control group receive 'standard' treatment, i.e. they develop spreadsheet formulae using the constructs and syntax in a spreadsheet application, such as Excel. The treatment group receive the novel approach, this allows relative comparison between the control and treatment groups.

## 2.4 Sampling

This sampling for this experiment is a cluster random sample as described by Shadish *et al.* (2002) and Saunders *et al.* (2007). Cluster sampling identifies a suitable cluster of participants and then randomly selects from within that group.

Considering similar development experiments in spreadsheets (Hicks and Panko 1995, Javrin and Morrison 1996, Panko and Halverson 2001), postgraduate Masters students were selected as an appropriate cluster.

Selection within the cluster was random, participants were not divided upon ability or any other basis.

Participants were invited to attend a session arranged for the experiment. Upon arriving participants were divided into two groups, the control and treatment groups. The appropriate materials for each group were distributed and the experiment began.

## 2.5 Research materials

The research materials for this experiment comprise two different packs handed to the participants.

Both packs contained a questionnaire gathering information such as age, sex, experience, number of years using spreadsheets, and a personal rating of their skill. This questionnaire was completed first, before the participants started the tasks. The point of this questionnaire is to gather demographic information and to determine the experience of spreadsheet use for a participant.

Once questionnaire 1 was completed, the participants started the tasks for the group they were assigned to (control or treatment). The scenarios contained in tasks for the participants, regardless of group, were identical. The manner in which the groups completed the tasks differed, the control group produced formulae in a spreadsheet using the syntax and functionality of the application (Microsoft Excel). The treatment group produced example attribute classifications for each task.

After completing the tasks as best they could, the final questionnaire, questionnaire 2, was completed. This questionnaire gathered information on the participant's perception of their own performance, i.e. they were asked how difficult they felt each task was and then asked to indicate how confident they were that the provided answers were correct.

## 2.6 Experiment tasks

The five tasks for the experiment were identical, the method of completing them varied for each group. The control group submitted answers created using Microsoft Excel, the treatment group submitted attribute classifications written on paper.

The experiment tasks were designed to be progressively more difficult, requiring progressively more complex answers from both groups.

**2.7 The tasks**

The tasks given to the control and treatment group were identical, the method in which they answered varied.

For example, in the control group task 1 was to create a formula that could give a grade (Pass or Fail) based upon a single mark (Exam mark). The formula was required to distinguish between pass and fail, where fail was < 40 and pass was >= 40.

For the same task, task 1, the treatment group were required to give attribute classifications (examples) for every classification in the problem. The two classifications are pass and fail, the participants therefore had to submit an attribute classification of pass and fail.

The tasks were also designed to be progressively more complex. For example task one uses one value (exam mark), 2 classifications (pass and fail) and two parameters for those classes (<40 Fail, >= 40 Pass).

In contrast, task 5 uses 2 values (exam and coursework mark), 4 classifications (fail, pass, merit and distinction) 4 parameters (< 40 fail, >= 40 pass, >= 55 merit and >= 70 distinction) and 1 conditional rule (Both exam and coursework values must fall in same class to award that class).

**2.8 Marking the control group**

Determining the mark of participants in the control group was based upon whether the answer provided was a valid formula in excel and whether the formula satisfied the specification in the task. If the formula fulfilled both criteria, it was deemed as correct, otherwise it is incorrect.

For incorrect formula, degrees of incorrectness were measured by counting the number of errors made in the submission. Errors can either be Mechanical, Logic or Omission, see Panko (1998) for a definition of these error types.

Once the number of errors was totalled, the submission was given a classification. The classifications were as follows: 0 errors = 5, 1 error = 4, 2-3 errors =3, 4 or more = 2, No attempt = 1.

These above classifications are used in the confidence calculation only, the other statistics are generated from dichotomous data.

**2.9 Marking the treatment group**

Determining the mark of the participants in the treatment group was based upon the whether the attribute classifications were valid and whether the attribute classifications provided satisfied the specification of the problem.

For incorrect attribute classifications, the number of errors per task was totalled and then given a classification. The classifications were as follows: 0 errors = 5, 1 error = 4, 2-3 errors =3, 4 or more = 2, No attempt = 1.

These above classifications are used in the confidence calculation only, the other statistics are generated from dichotomous data.

## *3.0 Summary statistics from experimentation*

In this section performance indicators are compared between the treatment and control groups. This indicates the usefulness of example giving in comparison to spreadsheet modelling.

### 3.1 Accuracy

By comparing accuracy results gained from both the treatment and control groups, it is evident that the treatment group were more accurate than the control group. See Figure 4

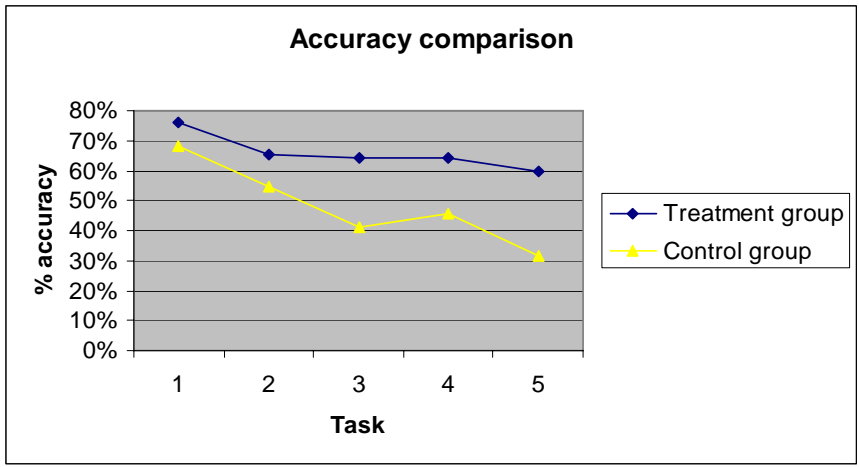

**Figure 3 Relative accuracy between Control and Treatment groups**

As can be seen, the treatment task accuracy ranges between 78 and 60 percent, the control group accuracy ranges between 66 and 30 percent. So comparatively, producing examples is more accurate than producing formulae.

### 3.2 Confidence

The confidence calculation indicates whether the group were perfectly calibrated, over or under confident. The formula for overconfidence is given in Figure 8 below.

$$Confidence\ ratio = \frac{Ratio\ percieved\ error\ rate}{Actual\ error\ rate}$$

**Figure 4 Confidence ratio calculation (Thorne *et al.* 2004)**

Further details of this calculation are contained in Thorne *et al.* (2004)

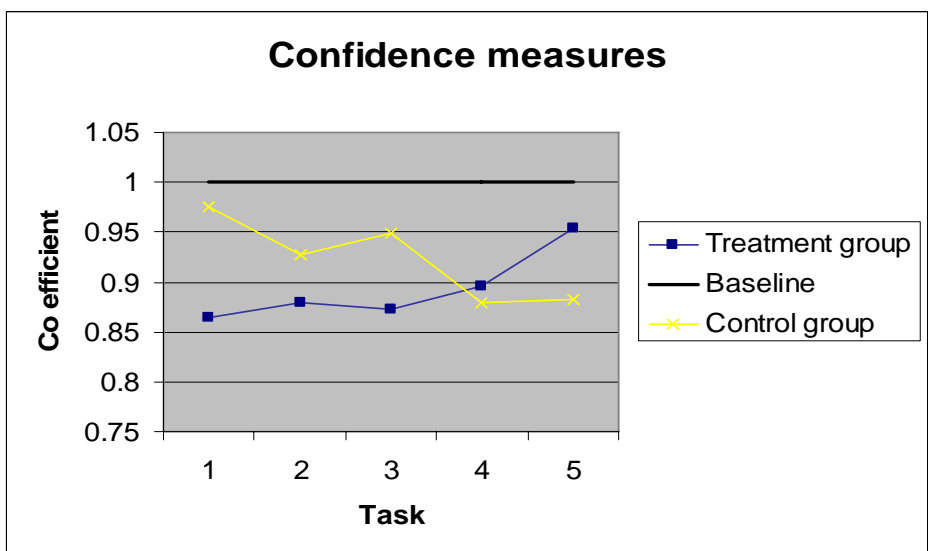

**Figure 5 Confidence in Treatment and Control groups**

The baseline on the graph shows the division between over and under confidence, a value of less than 1 indicates under confidence, over 1 indicates overconfidence. A value of 1 exactly indicates perfect calibration between expected outcome and performance.

As can be seen, both groups were under confident in their work. This is an unusual finding since the literature indicates that spreadsheet developers are usually overconfident (Panko, 2003).

Although the data in figure 9 shows that both groups were mostly under confident, there are some distinguishing features between them.

The treatment group's data points are less erratic than the control group, indicating a more consistent approach to evaluating their performance. This erratic grouping is clearer if perceived difficulty (how difficult was this task?) and Perceived completeness (did you complete the task successfully?) are mapped against each other, see figure 6.

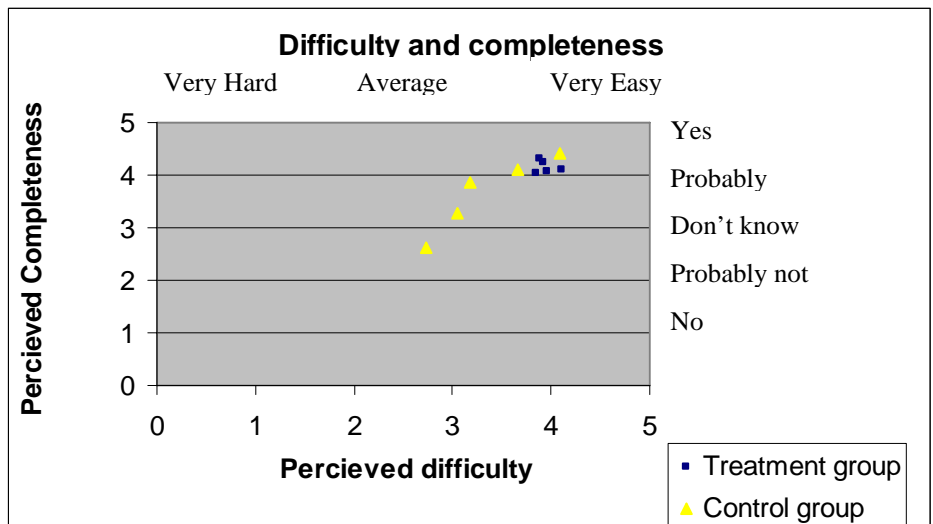

**Figure 6 Difficulty and completeness**

In figure 6, the treatment group's data points are bunched together, suggesting the values are similar. The values are responses to difficulty and completeness questions, this suggests that the treatment group found the task's difficulty and perceived completeness didn't change as the tasks progressed. In figure 6, the data points read right to left as tasks 1 to 5.

The control groups are more dispersed, indicating that the values change as the tasks progress, i.e. as the tasks progressed they were harder and perceived to be less complete.

## *4.0 Testing for statistical significance in the results*

### 4.1 Introduction

The raw data for both experiments, when graphed, allows conclusions to be drawn based up some basic statistics such as the mean value. Whilst this serves a purpose, it does not tell us if the results are statistically significant.

In order to see if the results are statistically significant a number of significance tests have been applied to the accuracy data. For example, the Chi squared test is used to determine if the differences in accuracy are statistically significant in the control and treatment groups. One can then determine if the increased accuracy observed in the treatment group was due to the treatment or not.

## 4.2 Chi squared test on accuracy data

The Chi squared test determines if the differences in accuracy for the treatment and control groups are due to the treatment and not chance. Once calculated, chi squared indicates if the "null hypothesis" should be accepted or rejected. The null hypothesis is usually the opposite of what the researcher wants to find, i.e. the null hypothesis is "There is no difference between the groups".

The raw data consists of 1's and 0's, the tasks were either correct (1) or incorrect (0). This characteristic of the data allows us to use the chi squared statistic in figure 7.

$$X^2 = \sum \frac{(\text{Observed frequencies - Expected frequencies})^2}{\text{Expected frequencies}}$$
$$= \sum \frac{(Fo - Fe)^2}{Fe}$$

**Figure 7 Chi squared statistic**

In cases where the sample size is small, Fisher's Exact test can be used to complement or replace the chi squared test (Fisher, 1922).

## 4.3 Fisher's exact test on accuracy data

Fisher's exact test determines the probability of the scenario being tested, or one more extreme, occurring. For clarity the test determines the probability of the same scenario or a more favourable one arising. Fisher's is applied when sample sizes are *small*, how small is unclear. Some cite less than 30 participants overall, some cite that less than 10 in a cell and some cite less than 4 in cell.

## 4.4 Summary of chi squared and Fisher's exact statistics

The combined results obtained from chi squared and Fisher's exact are contained in table 1 below.

|  | Chi squared test | Fisher's exact |
|---|---|---|
| Task 1 | 1.396<br>0.5 < P < 0.01<br>Accept Null | 0.205<br>80% |
| Task 2 | 0.673<br>0.5 < P < 0.01<br>Accept Null | 0.301<br>70% |
| Task 3 | 2.03<br>0.5 < P < 0.01<br>Accept Null | 0.128<br>88% |
| Task 4 | 2.03<br>0.5 < P < 0.01<br>Accept Null | 0.128<br>88% |
| Task 5 | 4.22<br>0.05 <P< 0.02<br>Reject null at 95% level. | 0.038<br>96% |

**Table 1 Combined Chi squared and Fisher's exact statistics**

The data in table 1 and the data graphed in figure 8, show that for both Chi squared and Fisher's exact, tasks 1 to 4 are not statistically significant, assuming that 95% is the minimum level of significance.

However, both show on task 5 statistical significance which therefore rejects the null hypothesis on that test. We can conclude that for task 5 the observed difference in accuracy was due to the treatment not chance.

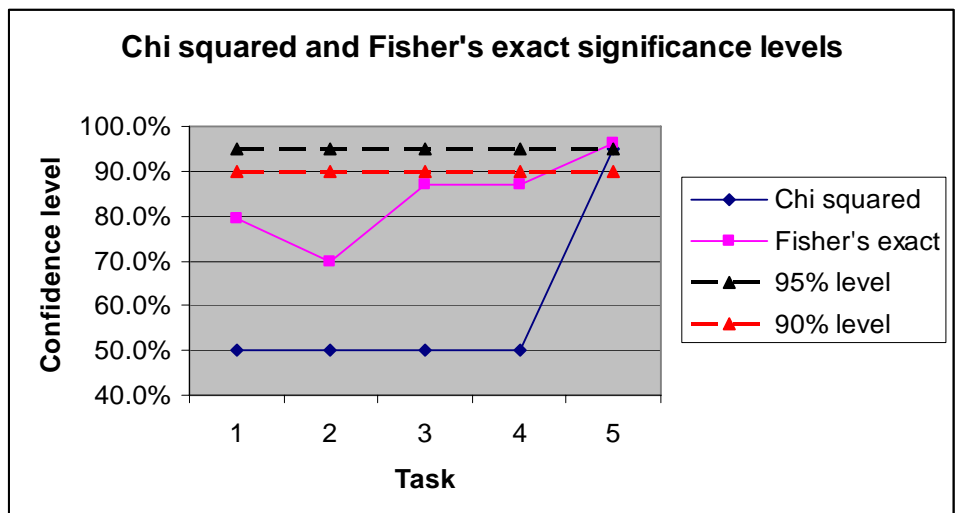

**Figure 8 Chi squared and Fisher's exact significance levels**

Since the tasks were designed to be progressively more difficult, one could interpret the results to show that the treatment is only effective in sufficiently complex scenarios.

Using Cochran's Q test determines if the difficulty between tasks was statistically significant

## 4.5 Cochran's Q test on difficulty

Cochran's Q test allows us to test if the difficulty between all five tasks in a particular group was significantly different. The test therefore has to be performed on both the control and treatment group. The formula for Cochran's Q test is given in figure 13.

$$Q = \sum_{j=1}^{c} \left( \frac{O_j - \frac{N}{c}}{\sqrt{\sum_{c=1}^{c} \frac{R_i}{c}\left(1 - \frac{R_i}{c}\right)}} \right)^2 = c(c-1)\frac{\sum_{j=1}^{c}(O_j - \frac{N}{c})^2}{\sum R_i(c - R_i)},$$

**Figure 9 Cochran's Q**

### 4.5.1 Cochran's Q for the Control group

The calculation for Cochran's Q statistic in the control group is as follows:

5 * 4 * (16 + 4 + 1 + 1 + 16)
= 760 / (270 – 194)
=10.00

DOF = 4
0.05 <P< 0.02

This shows that there is a significant difference in difficulty between tasks for the control group, we reject the null hypothesis at the 95% level.

### 4.5.2 Cochran's Q for the Treatment group

The calculation for Cochran's Q statistic for the treatment group is as follows:

5 * 4 * (10.24 + 0.04 + 0.64 + 0.64 + 3.24)
=296/(390-364)
=11.386

DOF = 4
Look up on Chi Squared table
0.05 <P< 0.02

This shows that there is a significant difference in difficulty between tasks for the treatment group, we reject the null hypothesis at the 95% level.

### 4.5.3 Conclusions on Cochran's Q test

The calculations of Cochran's Q test show that at the 95% confidence level, the null hypothesis, all the tasks are the same difficulty, for both the control and the treatment group is rejected. Therefore we can conclude that there is significant difference in difficulty between tasks.

This supports the theory that as the difficulty increases, the treatment effect becomes significant.

However, tasks 3 and 4 both show the same result for chi squared and Fisher's exact, see table 14. This might suggest that these two tasks were of similar difficulty based on the results.

In order to establish if this is the case, we must compare the two sets of data for the control and treatment group to see if there is statistical significance between them. One method to compare two data sets for difference in difficulty is McNemar's test on difficulty (McNemar, 1947).

## 4.6 McNemar's test on difficulty

The McNemar's statistic allows us to test for significant difference in difficulty between the two groups, in this case the results for task 3 and 4.
The test is $X^2$ using 1 DOF, see figure 10 for the equation.
$$X^2 = (b - c)^2/(b + c). \quad (1)$$

**Figure 10 McNemar's test on difficulty**

**McNemar's Calculations:**

$M = (3-3)^2 / (3+3) = 0/6 = 0$ **(Control Group)**

We therefore accept the null hypothesis, there is no difference between the two groups, i.e. there is no significant difference in difficulty between tasks 3 and 4 for the control group.

$M = (2-2)^2 / (2+2) = 0/4 = 0$ **(Treatment group)**

We therefore accept the null hypothesis, there is no difference between the two groups, i.e. there is no significant difference in difficulty between tasks 3 and 4 for the treatment group.

## 4.9 Conclusions on significance testing

The chi squared and Fisher's tests indicate that in both the control and treatment groups, for tasks 1 to 4, there is no statistically significant difference in accuracy.

However, both chi squared and Fisher's indicate that for task 5, in both control and treatment groups ,the observed increase in accuracy is statistically significant. i.e. the difference in accuracy is due to the treatment and not chance, ergo giving examples in task 5 is more accurate than producing the equivalent formula.
Cochran's Q test indicates that between all five tasks, there is a significant difference in difficulty. McNemar's test on the observed accuracy in tasks 3 and 4, which have the same values, demonstrates that there is no significant difference in difficulty between the tasks.

One possible explanation is that during the design of the materials, i.e. the tasks were not sufficiently different to yield a significant change in difficulty, hence the same accuracy values.

To conclude, there is a relationship between difficulty and statistically significant accuracy for the treatment. The results suggest that if the task or problem is sufficiently difficult, there is a statistically significant accuracy advantage in using the treatment over the control.

## *5.0 Conclusions*

The conclusions of the experimental comparison between the Treatment group, i.e. giving examples and control group, i.e. producing formulae

## 5.1 Experimental Conclusions

1. The treatment group (giving examples) were considerably more accurate than the control group (producing formulae), see figure 4. Accuracy in task 5 only was task to be statistically significant, see table 2 and figure 14.
2. Both the treatment group (giving examples) and the control group (producing formulae) were consistently under confident, see figure 10.
3. Both groups found the tasks progressively more difficult as Cochran's Q test indicated, except tasks 3 and 4 which showed no significance of this type, see section 3.5.7.

## 5.2 Limitations

Limitations to this experimental study include both general criticisms of experimental work and specific conditions that relate to the experiment. Also so criticism could be made of the statistical significance tests due to the way that they are marked.

### 5.2.1 Criticisms of the experiment

Firstly, the sample of participants is from an academic environment, experimentation with participants from a non academic environment would provide a broader view of the usefulness of this method.

Although there was no time limit imposed on the participants to complete the tasks, participants were not permitted to take the materials away from the venue. Some might argue that this imposes a time pressure on the participants and that in reality they are more likely to complete the tasks over a longer time period.

However, to keep control of the experimental conditions one must insist that participants stay in the arranged venue until they have completed. Allowing them to remove and complete materials at another venue may allow collusion and thus the integrity of the experiment would be compromised.

It could be argued that the sampling approach taken in this experiment is not truly random. A clustered random approach was taken, i.e. a cluster of individuals were targeted and then randomly assigned to either the treatment or control group.

### 5.2.3 Criticisms of the significance testing

The significance tests show that only task 5 is statistically significant. The Cochran's Q statistic shows that the difficulty difference between the tasks is statistically significant.

The tasks were designed to be progressively more difficult. The conclusion is therefore that the treatment effect is only statistically significant in sufficiently difficult tasks.

The statistics generated from the raw data are sensitive to the marking applied to the answers provided to each question. The answers were dichotomous, i.e. attempts were either correct or incorrect. In both the control and treatment group this mark was based upon whether the solution provided was a valid solution that covered the specification of the task.

If the method used to mark the answers provided for each task differed, one would expect to see a change in the statistics. If the statistics were calculated data that had been processed according to an invented marking criteria, the sensitivity of the statistics would be greater.

However, since all of the statistics were strictly marked in a dichotomous fashion, this sensitivity is not a limiting factor in this research.

## *5.3 Conclusion on the novel approach*

The results of the experiment demonstrate that giving examples is more accurate, easier and less prone to overconfidence than creating formula. It is therefore feasible to use "giving examples" as the basis for a modelling method.